\documentclass[aps,prd, final, superscriptaddress,nofootinbib]{revtex4}

\usepackage{hyperref}
\usepackage{graphicx}
\usepackage{amsmath}
\usepackage{amssymb}
\usepackage{xspace}
\usepackage{multirow}

\newcommand{\eq}[1]{eq.~\eqref{eq:#1}}

\renewcommand{\sec}[1]{sec.~\ref{sec:#1}}
\newcommand{\secs}[2]{secs.~\ref{sec:#1} and \ref{sec:#2}}

\newcommand{\app}[1]{app.~\ref{app:#1}} 
\newcommand{\fig}[1]{fig.~\ref{fig:#1}}
\newcommand{\figs}[2]{figs.~\ref{fig:#1} and \ref{fig:#2}}
\newcommand{\eg}{{\it e.g.~}}

\newcommand{\ord}[1]{{\mathcal O}(#1)}
\newcommand{\ORd}[1]{{\mathcal O}\Bigl(#1\Bigr)}

\newcommand{\nn}{\nonumber}

\newcommand{\df}{\mathrm{d}}
\newcommand{\img}{\mathrm{i}}
\newcommand{\sdt}{\!\cdot\!}

\newcommand{\al}{\alpha}

\newcommand{\ga}{\gamma}

\newcommand{\de}{\delta}

\newcommand{\si}{\sigma}

\newcommand{\cJ}{{\mathcal J}}

\newcommand{\cO}{{\mathcal O}}

\newcommand{\GeV}{\,\mathrm{GeV}}

\newcommand{\lqcd}{\Lambda_\mathrm{QCD}}

\newcommand{\cut}{\mathrm{c}}

\newcommand{\zero}{{(0)}}
\newcommand{\one}{{(1)}}

\newcommand{\Pythia}{\textsc{Pythia}\xspace}

\begin{document}


\title{Fragmentation with a Cut on Thrust: \\ Predictions for B-factories}

\author{Ambar Jain}
\affiliation{Department of Physics, Carnegie Mellon University, Pittsburgh, PA~15213, U.S.A.}\affiliation{Physics Department, Indian Institute of Science Education and Research, Bhopal, M.P. 462023, India}
\author{Massimiliano Procura}
\affiliation{Albert Einstein Center for Fundamental Physics, Institute for Theoretical Physics,University of Bern, CH-3012 Bern, Switzerland}
\author{Brian Shotwell}
\affiliation{Department of Physics, University of California at San Diego, 
La Jolla, CA 92093, U.S.A.}
\author{Wouter J.~Waalewijn}
\affiliation{Department of Physics, University of California at San Diego, 
La Jolla, CA 92093, U.S.A.}


\begin{abstract}
When high-energy single-hadron production takes place inside an identified jet, there are important
correlations between the fragmentation and phase-space cuts.
For example, when one-hadron yields are measured in on-resonance B-factory data, a cut on the thrust event shape $T$ is required to remove the large $b$-quark contribution. This leads to a dijet final state restriction for the light-quark fragmentation process. Here we complete our analysis of unpolarized fragmentation of (light) quarks and gluons to a light hadron $h$ with energy fraction $z$ in $e^+ e^- \to \text{dijet} + h$ at the center-of-mass energy $Q=10.58$ GeV. In addition to the next-to-next-to-leading order resummation of logarithms of $1-T$, we include the next-to-leading order (NLO) nonsingular $\ord{1-T}$ contribution to the cross section, the resummation of threshold logarithms of $1-z$, and the leading nonperturbative contribution to the soft function. Our results for the correlations between fragmentation and the thrust cut are presented in a way that can be directly tested against B-factory data. These correlations are also observed in \Pythia, but are surprisingly smaller at NLO.
\end{abstract}



\maketitle

\section{Introduction}
\label{sec:intro}

Hard QCD interactions give rise to highly virtual partons that evolve by emitting radiation, until they hadronize into the nonperturbative states we observe. The QCD process underlying hadron ``fragmentation" has not been as well quantified as jet production or deep-inelastic scattering (DIS), making this an active area of research. Understanding parton fragmentation clearly extends the set of reactions that can be handled within a perturbative QCD approach. Applications include, for example, the investigation of the spin structure of the nucleon in semi-inclusive DIS~\cite{Airapetian:2004zf,Alekseev:2010ub} and the study of hadron production at high $p_T$ in $p p$ collisions~\cite{Agakishiev:2011dc,Adare:2007dg} which is crucial to determine more accurately the relative suppression of hadron spectra (jet quenching) seen in heavy-ion collisions. In general, for many such applications additional cuts on the hadronic final state may be required to reduce the background from other processes, to help identify underlying partonic structures (see \eg refs.~\cite{Krohn:2012fg,Waalewijn:2012sv}). 

A key observation is that the fragmentation process involves physics at well-separated energy scales. It has been shown that high-energy processes with an observed hadron in the final state can be described by factorizing the short-distance (partonic) physics associated with the hard scale $Q$, which is perturbatively calculable in QCD, from (universal) nonperturbative long-distance contributions, see \eg ref.~\cite{Collins:1989gx}. 
At leading power in $\lqcd/Q$, the (unpolarized) fragmentation functions (FFs) $D_i^h(z,\mu)$~\cite{Collins:1981uk, Collins:1981uw} encode the information on the nonperturbative transition from an energetic parton $i=\{g,u,\bar u, d, \dots\}$ to a hadron $h$, which carries a fraction $z$ of its energy \footnote{Due to perturbative radiation, the momentum fraction in the fragmentation function is not the same as the experimentally measured one beyond the leading order [see \eg \eq{factfirst}].}, plus a remainder $X$. The knowledge of both perturbative and nonperturbative ingredients in factorization theorems is crucial to obtain theoretical predictions. Unpolarized FFs, for example, serve as an input to extract the flavor-separated helicity distributions from spin asymmetries observed in polarized semi-inclusive processes~\cite{deFlorian:2009vb,Leader:2010rb}. 

Parameterizations for the spin-averaged FFs have been constrained by fitting to data for single-inclusive charged hadron production in $e^+ e^-$ at next-to-leading order (NLO) in perturbation theory~\cite{Albino:2005me,Albino:2005mv,Hirai:2007cx}. More recently, global analyses have been performed to incorporate also semi-inclusive deep-inelastic scattering and/or $p p$, $p \bar{p}$ data from HERA, RHIC and the Tevatron~\cite{deFlorian:2007aj,deFlorian:2007hc,Albino:2008fy}. To illustrate the current level of accuracy, the dominant $D_u^{\pi^+}\!(z,\mu=m_Z)$ is determined with uncertainties at the 10\% level for $z \gtrsim 0.5$~\cite{Hirai:2007cx}. The FFs of the gluon and the non-valence quarks are known even less accurately. 

The inclusion of high-statistics B-factory data in fragmentation function analyses will significantly improve the precision with which these are extracted. Most of the $e^+e^-$ data is so far at $Q=m_Z$ and phenomenological input at a different scale will give access to the gluon FF via mixing with the quark FFs in the renormalization group evolution. The Belle collaboration has very recently analyzed hadron multiplicities in off-resonance data~\cite{Leitgab:2013qh}, which can be used to extract fragmentation functions. If on-resonance data were also analyzed, a cut on thrust $T$ would be needed to remove the large $b$-quark background. This was carried out in ref.~\cite{Seidl:2008xc} for the Collins effect and has been considered to study unpolarized fragmentation as well~\cite{Belle:MGP}. The thrust event shape variable is defined as~\cite{Farhi:1977sg}
\begin{equation}
  T = \max{}_{\hat t}\; \frac{\sum_i |\hat t \sdt {\vec p}_i|}{\sum_i |{\vec p}_i|} 
  \,,
\end{equation}
where the sum is over all final-state particles. More precisely, it is found that a thrust cut of $\tau=1-T < 0.2$ removes $98\%$ of the $B$ data leaving the thrust distribution dominated by the fragmentation of light ($u$, $d$, $s$) and charmed quark pairs~\cite{Seidl:2008xc}. With this cut the hadronic final state is given by back-to-back jets. 

\begin{figure}[t]
\centering
\includegraphics[width=0.48\textwidth]{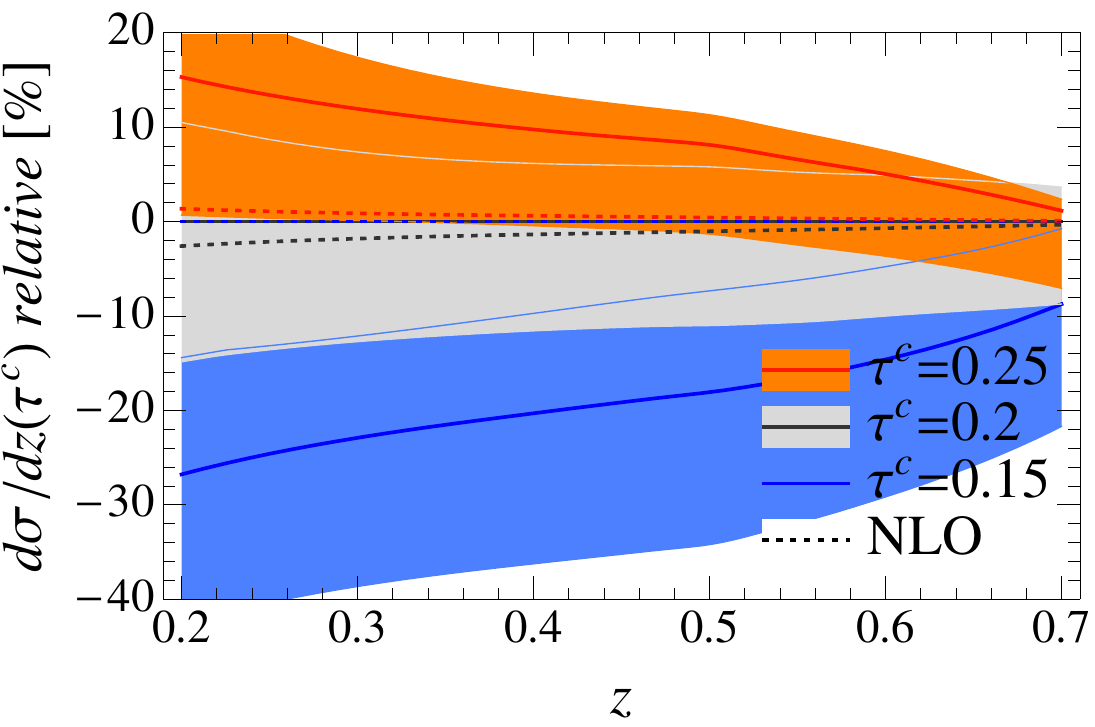} \quad
\includegraphics[width=0.48\textwidth]{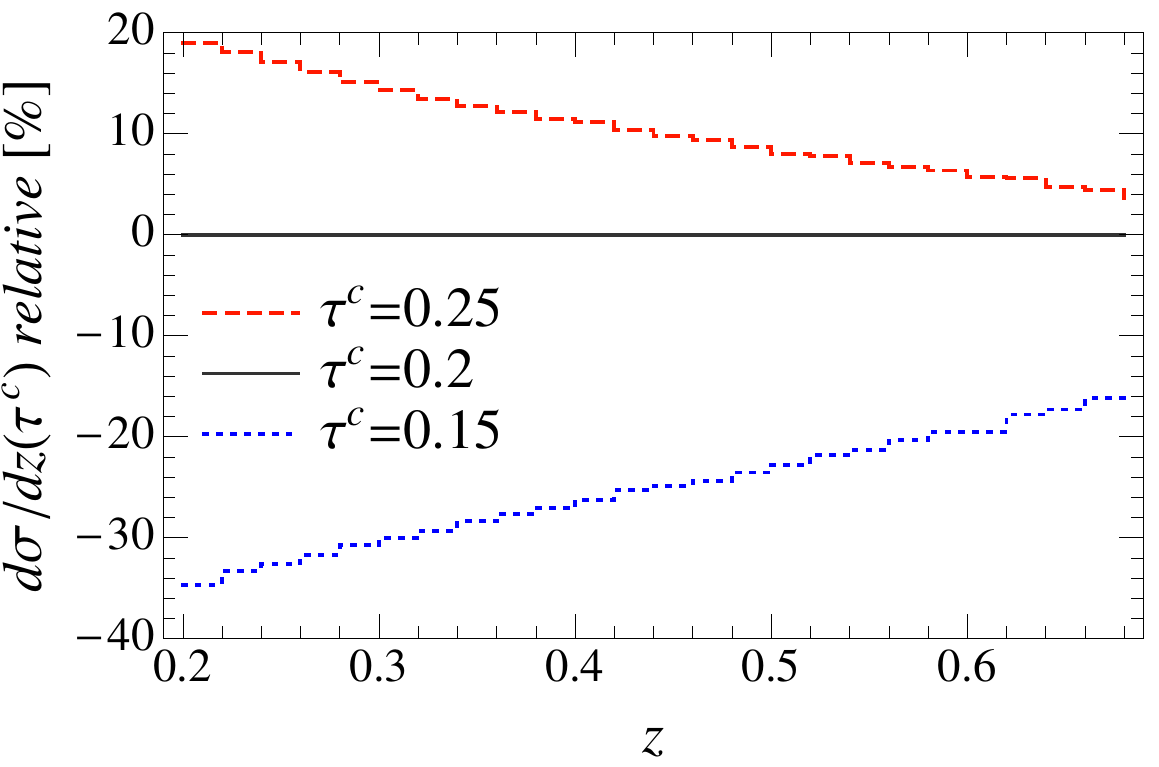}
\caption{Correlations between the thrust cut $\tau \leq \tau^\cut$ and the observed momentum fraction $z$ in the cross section of $e^+ e^- \to \text{dijet} + \pi^+$ for $Q=10.58$ GeV. Curves and bands are plotted relative to the case $\tau^\cut=0.2$. A cut on thrust changes the shape in $z$. Left panel: our NNLL+NLO results with perturbative uncertainty bands. The result at NLO (dotted lines) contains negligible $\tau^\cut$-$z$ correlations. As input we use the HKNS fragmentation functions $D_i^{\pi^+}$~\cite{Hirai:2007cx} at NLO. Right panel: the same observable using \Pythia.}
\label{fig:moneyplot}
\end{figure}

In refs.~\cite{Procura:2009vm, Jain:2011xz, Jain:2011iu, Procura:2011aq} we developed the theoretical framework to study the fragmentation of a light parton into a light hadron $h$ inside an identified jet. With ``identified" we refer to jets determined either through an event shape like thrust~\cite{Procura:2009vm, Jain:2011xz} or via a jet algorithm~\cite{Procura:2011aq,Waalewijn:2012sv}. Our analysis provides the theoretical tools for a more exclusive study of fragmentation than is currently done, which is suitable for processes where additional cuts on the hadronic final states are imposed, such as the Belle proposal in ref.~\cite{Seidl:2008xc}. 
We focus on the spin-averaged case since our main purpose here is to provide the tools for a straightforward test of our framework against high-statistics on-resonance B-factory data. We expect that a generalization of our formalism to study the azimuthal asymmetric distribution of hadrons inside an identified high energy jet can lead to interesting applications in the context of spin-dependent fragmentation~\cite{Collins:1993kq,Yuan:2007nd,D'Alesio:2010am}. 

Our framework allows us to calculate the non-trivial correlations between the thrust cut $\tau^\cut$ and the energy fraction $z$ in the observed cross section for $e^+ e^- \to \text{dijet} + h$. Due to these correlations, the thrust cut modifies the shape in $z$. As already shown in the preliminary study of ref.~\cite{Jain:2011xz}, where we resummed the logarithms $\al_s^n \ln^m \tau$ to next-to-next-to-leading logarithmic accuracy (NNLL), the effect is expected to be sizable and important for the analysis of data on the $\Upsilon(4S)$. Here we improve our study in ref.~\cite{Jain:2011xz} by including the NLO nonsingular contribution to the cross section, the resummation of threshold logarithms of $1-z$ and the leading nonperturbative correction to the thrust soft function, which are discussed in \sec{calc}. 

The correlations we obtain from our new analysis are shown in the left panel of \fig{moneyplot} for various thrust cuts $\tau^\cut$, including the theoretical (perturbative) uncertainties. As can be seen there, the effect of the thrust cut $\tau^\cut$ is much stronger in the small $z$ region than in the large $z$ region. This has a straightforward physical explanation: for larger values of $z$, most of the jet energy is carried by the observed hadron, causing the jet to be more collimated and less affected by the thrust cut. In particular, for $z \geq 1- \tau^\cut$, the cross section is unaffected by the thrust cut. On comparison with the right panel, we see that the effect we find is fully compatible with the outcome of the \Pythia event generator~\cite{Sjostrand:2006za,Sjostrand:2007gs} based on leading logarithmic parton showering and the Lund string fragmentation model \footnote{The results for other light hadrons, like the proton, are very similar to the ones shown here for $\pi^+$.}. We stress that the plot in the left panel has been obtained from our purely perturbative calculation, taking only the phenomenological parameterization of the universal $D_i^{\pi^+}$ in ref.~\cite{Hirai:2007cx} as input. There we also show the correlations that we obtain at NLO (dotted curves), which are much smaller. Presumably, the single additional emission at NLO is insufficient to reliably describe this doubly-differential cross section, compared to the multiple emissions in our resummed calculation and in \Pythia. 

The paper is organized as follows. In \sec{calc} we briefly review our theoretical framework, where we describe the updates of our analysis compared to our previous work in ref.~\cite{Jain:2011xz}. In \sec{num} we discuss the features of the $z$-spectrum and $\tau$-spectrum and illustrate the effect of the different ingredients of our calculation. 
In \sec{prop} we propose a first quantitative analysis of correlations between $z$ and $\tau^\cut$ which can be straightforwardly tested on Belle data. 
All the necessary theoretical input for this moment-space analysis is collected in \app{coeff}. App.~\ref{app:nonsingular} contains the details of the calculation of the NLO nonsingular contribution to the cross section. A detailed discussion of our choice of renormalization scales, as well as the scale variations used to estimate the perturbative uncertainties, is given in \app{scales}.

\section{Calculating correlations between thrust and $z$ in $e^+ e^- \to \text{dijet} + h$}
\label{sec:calc}

We focus on the case of spin-averaged fragmentation in $e^+ e^- \to \text{dijet} + h$ where one restricts to the dijet limit by a cut on thrust. Starting from the well-known factorization theorem for the inclusive measurement of thrust~\cite{Catani:1992ua,Korchemsky:1999kt,Fleming:2007qr,Schwartz:2007ib}, we obtained in refs.~\cite{Procura:2009vm, Jain:2011xz} a factorization theorem for the observed cross section at leading power in $\lqcd/Q$, which has this form: 
\begin{equation} \label{eq:factfirst}
  \frac{\df \si^h}{\df z}(\tau^\cut) =  \int_0^{\tau^\cut}\!\! \df \tau\,\frac{\df^2 \si^h}{\df \tau\,\df z} = 
\! \sum_{j=g,\, u, \,\bar{u},\, d \dots} \int_z^1 \frac{\df x}{x}\, D_j^h(x,\mu) \;C_j\Big(\tau^\cut, \frac{z}{x}, Q^2, \mu \Big)\, ,
 \end{equation}
where $Q$ is the center-of-mass energy. The $D_j^h(x)$ denote the standard, unpolarized fragmentation functions, which describe the hadronization $j \to h(x)$ at leading power. The index $j$ runs over all parton flavors, including the gluon. The coefficients $C_j$ are calculable in perturbation theory and contain double logarithms of $\tau^\cut$, $\al_s^n \ln^m \tau^\cut$ ($m\leq 2n$), and ``threshold" logarithms of $1-z$. Our setup enables us to resum all these large logarithms in the dijet region of the thrust distribution. This is necessary to achieve perturbative convergence with small uncertainties in $C_j$ when $\tau^\cut \lesssim 0.2$ and when $z \gtrsim 0.5$, as we will show in \sec{num}. The scale $\mu$ in \eq{factfirst} is arbitrary, but must be chosen to be the same in $C_j$ and $D_j^h$.

\begin{figure}[t]
\centering
\includegraphics[width=0.9\textwidth]{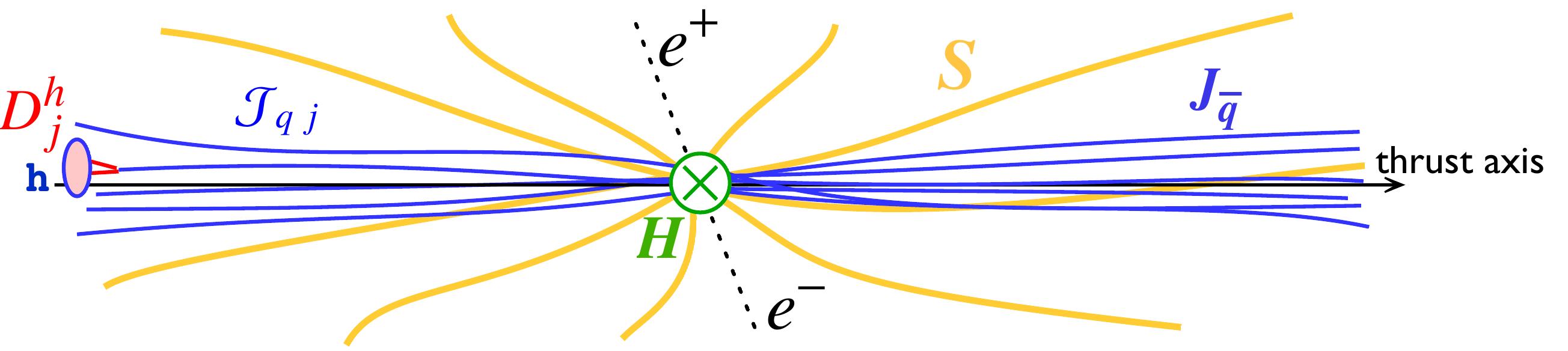}
\caption{A schematic of the various subprocesses in $e^+ e^- \to \text{dijet} + h$: The (green) vertex denotes the hard interaction $H$, the (blue/dark) jets are described by $J$ and $\cJ_{ij}$ (the latter for the jet in which the hadron is observed), the fragmentation $j \to h$ is described by the fragmentation function $D_j^h$ and the effects of the (yellow/light) soft radiation are contained in $S$.}
\label{fig:fact}
\end{figure}
We obtained the expression for $C_j$ by exploiting factorization in the framework of soft-collinear effective theory~\cite{Bauer:2000ew,Bauer:2000yr,Bauer:2001ct,Bauer:2001yt}. The underlying idea is that dynamics at well-separated scales combine incoherently, allowing one to factorize their contributions to the cross section. In the dijet region, the relevant subprocesses are illustrated in \fig{fact}.  
Specifically, we distinguish the scale $Q$ for the hard collision $e^+ e^- \to q \bar{q}$, the scale $\sqrt{\tau^\cut} Q$ for the jet production (showering), the scale $\tau^\cut Q$ associated with the soft radiation between the jets, and the scale $\lqcd$ at which hadronization takes place. The coefficients $C_j(\tau^\cut, Q^2, x/z,\mu)$ are given by the convolution of functions encoding the effects at these different perturbative scales plus a nonsingular piece which contains contributions that do not factor in this way. In detail, 
 \begin{align} \label{eq:Cj}
C_j\Big(\tau^\cut, Q^2, \frac{x}{z},\mu\Big) =&\sum_q \frac{\sigma_0^q}{2 (2 \pi)^3}\,  H(Q^2, \mu) \,
\int_0^{\tau^\cut}\!\! \df \tau \int\! \df s_a\, \df s_b \Big[\cJ_{q j} \Big(s_a, \frac{z}{x}, \mu \Big)\, J_{\bar{q}} (s_b,\mu) + J_q(s_a, \mu)\, \cJ_{\bar{q} j} \Big(s_b,\frac{z}{x},\mu \Big)\Big]
 \nn \\
 & \times Q\, S_\tau \Big(Q \tau - \frac{s_a + s_b}{Q},\mu\Big)  \,
\Big\{1+ \cO\Big[\frac{\lqcd^2}{(1-z)\tau^\cut Q^2},\Big(\frac{\lqcd}{\tau^\cut Q}\Big)^2\Big]\Big\}
  + C^{\rm ns}_j\Big(\tau^\cut, Q^2, \frac{x}{z},\mu\Big) \,.
\end{align}
Here $\sigma_0^q$ is the tree-level cross section for the electroweak process $e^+ e^- \to (\gamma\,, Z) \to q \bar q$, which depends on the quark flavor. $H$ is the hard function that encodes virtual corrections to the production of the $q \bar q$ pair at the hard scale $\mu_H \simeq Q$. It is given by the square of the Wilson coefficient in the matching of the quark current from QCD onto SCET:
 \begin{equation} \label{eq:hard}
 H(Q^2,\mu_H) =  \biggl \lvert 1 + \frac{\alpha_s(\mu_H)\,C_F}{4\pi} \biggl[-\ln^2 \Bigl(\frac{-Q^2-\img 0}{\mu_H^2}\Bigr) + 3 \ln \Bigl(\frac{-Q^2-\img 0}{\mu_H^2}\Bigr) - 8 + \frac{\pi^2}{6} \biggr] \biggr \rvert ^{\,2}\, .
 \end{equation}
In order to improve convergence, we resum the large $\pi^2$-terms that arise here by taking $\mu_H=-\img Q$~\cite{Parisi:1979xd, Sterman:1986aj, Magnea:1990zb, Eynck:2003fn}.

The collinear contributions to the invariant masses of the jets are denoted by $s_a$ and $s_b$, and combine with the contribution from soft radiation to give the thrust $\tau$. This leads to the convolution in \eq{Cj}. The perturbative coefficients $\cJ_{ij}$ describe the emissions from the parent parton building up the jet within which the hadron $h$ fragments. The invariant mass distribution of the jet in the hemisphere opposite to $h$ is described by the (perturbative) inclusive jet function $J_j$. Both functions are characterized by the scale  $\mu_J \simeq \sqrt{\tau}Q$. Since it is not known whether the observed hadron $h$ fragmented from the quark or the antiquark jet, in \eq{Cj} we sum over both possibilities. The soft function $S_\tau$ describes the contribution to thrust due to soft parton emissions. It is defined through the vacuum matrix element of eikonal Wilson lines and the corresponding soft scale is $\mu_S \simeq \tau Q$. 

In ref.~\cite{Jain:2011xz} we calculated the quark and gluon $\cJ_{ij}$'s at one loop (the results are partly contained also in ref.~\cite{Liu:2010ng}) and this enabled us to analyze the cross section in \eq{factfirst} up to NNLL since $H$ and $J_j$ are already known in the literature to the relevant accuracy. The $\cJ_{ij}$'s contain plus distributions in $s_i/\mu_J^2$, which lead to double logarithms after the integral over $s_i$. As pointed out in ref.~\cite{Procura:2011aq} the threshold logarithms in $1-z$ appearing in $\cJ_{ij}$ can simultaneously be resummed through an appropriate choice of $\mu_J$.

Finally, the coefficient  $C^{\rm ns}_j$ contains the terms of the NLO cross section that are not enhanced by logarithms of $\tau$, and is $\ord{\tau}$ suppressed. The calculation of the nonsingular contribution to $\df^2 \si^h/(\df \tau\, \df z)$ at NLO is discussed in detail in \app{nonsingular} and its numerical effects are shown in \sec{num}. 
Lastly, we stress that since fragmentation takes place inside a (hemisphere) jet of invariant mass of order $\tau^\cut Q^2$, the momentum fraction $z$ cannot be too small, to avoid contributions from soft hadrons that are not to be associated with the jet, which originates from an energetic (anti)quark. The hadron momentum must therefore be larger than the soft scale, implying $z \gtrsim \tau^\cut$. 

As we just discussed, the factorization in \eq{Cj} enables the resummation of logarithms of $\tau$ and $1-z$, as well as large $\pi^2$-terms in $H$, which is accomplished by evaluating $H$, $\cJ_{ij}$, $J_j$ and $S_\tau$ at their natural scales and using their respective RGEs to evolve them to a common scale $\mu$. We have made improvements in our choice of scales compared to our earlier work in ref.~\cite{Jain:2011xz}. Specifically, we resum threshold logarithms and take the kinematic upper bound $\tau \leq 1-z$ into account. A detailed discussion of our choice for the running scales is contained in \app{scales}.

We now comment on the various corrections in \eq{Cj}. There are corrections of order $\lqcd^2/[(1-z)\tau^\cut Q^2]$ associated with the $\cJ_{ij}$'s. These coefficients describe the emissions from the parent parton (here a quark) at large virtualities building up the jet in which the hadron is identified. The associated corrections get sizable if the jet invariant mass gets small. In our plots in \sec{num} we have included the leading nonperturbative contribution to $S_\tau$, according to the analysis in ref.~\cite{Abbate:2010xh}, which pushes the nonperturbative corrections to the soft function to order $\lqcd^2/(\tau^\cut Q)^2$, as indicated in \eq{Cj}. These nonperturbative corrections become large for small $\tau^\cut$, up to about $15\%$ for $\tau^\cut =0.15$. We therefore need $\tau^\cut$ to be not too small.

Quark and hadron masses are treated as negligible in our calculation, and we briefly comment on the size of the corresponding kinematic corrections. The corrections due to the hadron mass $m_h$ are of order $m_h^2/(z^2 Q^2)$ \cite{Albino:2005gd}. For light hadrons, like pions and kaons, they are indeed negligible in the region where our framework can be applied. In the plots of \sec{num}, we restrict $q$ in \eq{Cj} to be $u,d,s$. The correction due to the charm mass is expected to be of order $m_c^2/(\tau^\cut Q^2)$ when charm is a valence quark, which is less significant than the other corrections to \eq{Cj}. If the hadron $h$ does not contain a charm valence quark, the only contribution to \eq{factfirst} that does not get further suppressed by the smallness of the corresponding FF, is the one involving $D_g^h$.

\section{Numerical analysis for $e^+ e^- \to \text{dijet} + \pi^+$}
\label{sec:num}

Here we illustrate our key results for the cross section in $e^+ e^- \to \text{dijet} + \pi^+$, given in \eq{factfirst}. As input we use the HKNS fragmentation functions $D_q^{\pi^+}$ and $D_g^{\pi^+}$~\cite{Hirai:2007cx} at NLO \footnote{We have verified that using an alternative FF set at NLO (either DSS~\cite{deFlorian:2007aj} or AKK08~\cite{Albino:2008fy}) does not alter the main results of this paper.}. To be consistent with ref.~\cite{Hirai:2007cx}, we set $\alpha_s(\mu=m_Z) = 0.125$ with two-loop running, matching continuously across the quark thresholds \footnote{In the next sections, which do not involve HKNS FFs, we will use a more realistic value for $\alpha_s$.}. In our analysis we do not include the effects of the uncertainties associated with $\al_s(m_Z)$ or with the FF parameters. Therefore we stress that the bands shown in our plots correspond only to the perturbative uncertainties obtained from scale variations, as explained in \app{scales}.

We have made the following improvements with respect to the plots shown in ref.~\cite{Jain:2011xz}:
\begin{itemize}
\item[(a)] We include the NLO nonsingular contribution, which we calculate in \app{nonsingular}. This small $\ord{\tau}$ correction plays a role for $\tau^\cut \gtrsim 0.2$.
\item[(b)] We resum the threshold logarithms of $1-z$ according to our ref.~\cite{Procura:2011aq}, where we introduced a joint resummation of these logarithms and those in $\tau^\cut$ (suitable for values of $\tau^\cut$ in the range of interest for a test with Belle data). As described in \app{scales}, this is accomplished through the choice of the renormalization scale for $\cJ_{ij}$. These effects are expected to play a role for $z \gtrsim 0.5$.
\item[(c)] The leading $\lqcd/(Q \tau)$ power correction in $\df \sigma/ \df \tau$ is known to be a shift~\cite{Manohar:1994kq,Webber:1994cp,Korchemsky:1994is,Dokshitzer:1995zt}. This may be seen by performing the operator product expansion on $S_\tau$~\cite{Abbate:2010xh}
\begin{align}
S_\tau(k)=S_\tau^{\rm part}(k)-2 \bar\Omega_1\frac{\df S_\tau^{\rm part}(k)}{\df k} + \ORd{\frac{\lqcd^2}{k^3}}
= S_\tau^{\rm part}(k-2\bar\Omega_1) + \ORd{\frac{\lqcd^2}{k^3}}
\,,\end{align}
where $S_\tau^{\rm part}$ is the partonic soft function and $\bar{\Omega}_1$ is a nonperturbative matrix element in the $\overline{\rm MS}$ scheme. Inserting the second expression in \eq{Cj}, this leads to a shift in the thrust distribution $\tau^\cut \to \tau^\cut-2 \bar{\Omega}_1/Q$. Its effect is most prominent at small $\tau^\cut$. From the analysis in ref.~\cite{Abbate:2010xh}, $\bar{\Omega}_1=0.252\,{\rm GeV}$. 
\end{itemize}
\begin{figure}[t]
\centering
\includegraphics[width=0.48\textwidth]{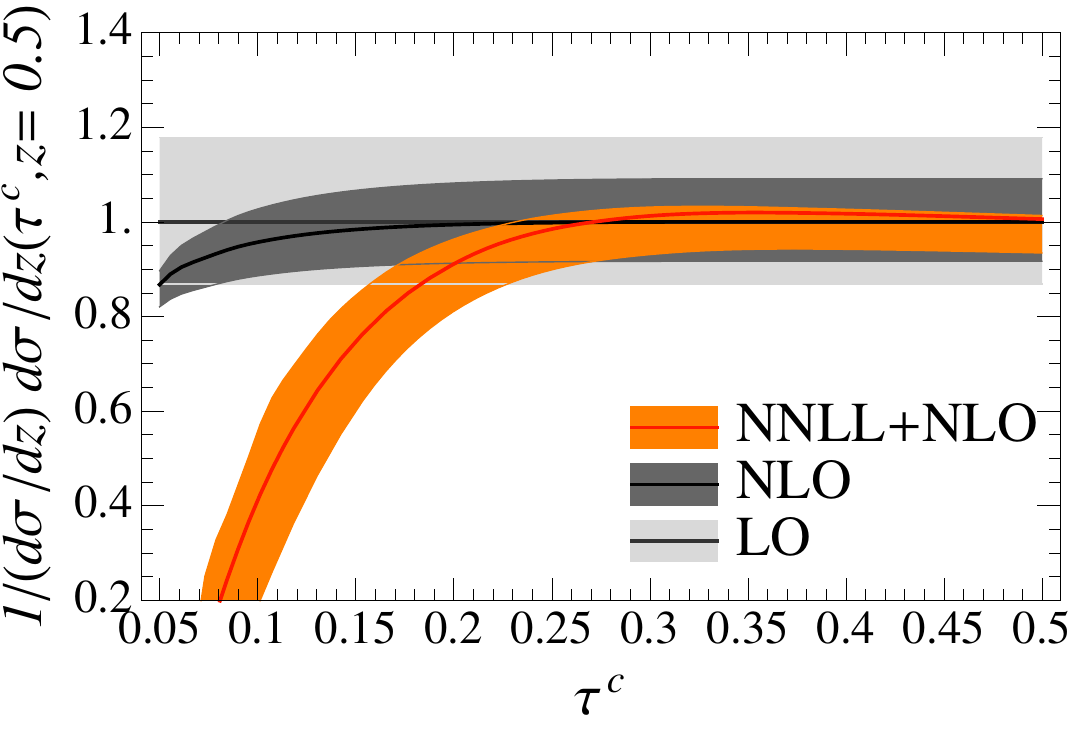}
\caption{The effect of the thrust cut in $e^+ e^- \to \text{dijet} + \pi^+$ for $Q=10.58$ GeV and $z=0.5$ at LO, NLO and NNLL+NLO. The curves are normalized at $\tau^\cut=0.5$. The bands show the perturbative uncertainties obtained from scale variations discussed in \app{scales}.}
\label{fig:si_tau}
\end{figure}
\begin{figure}
\centering
\includegraphics[width=0.475\textwidth]{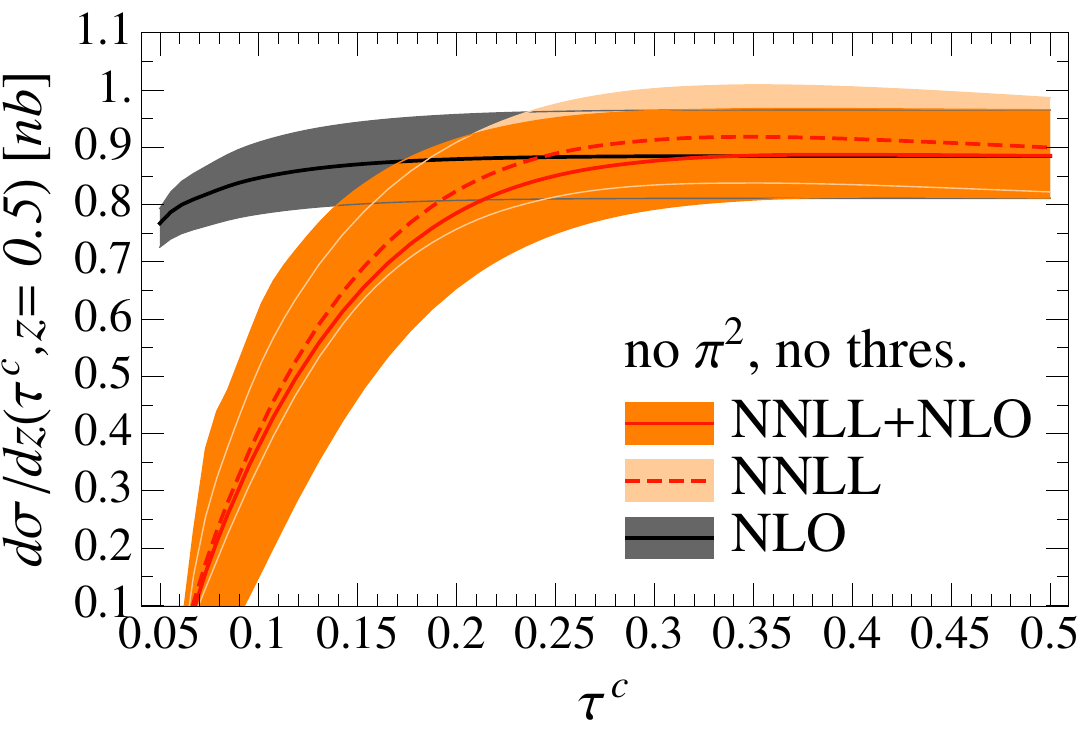} \quad
\includegraphics[width=0.483\textwidth]{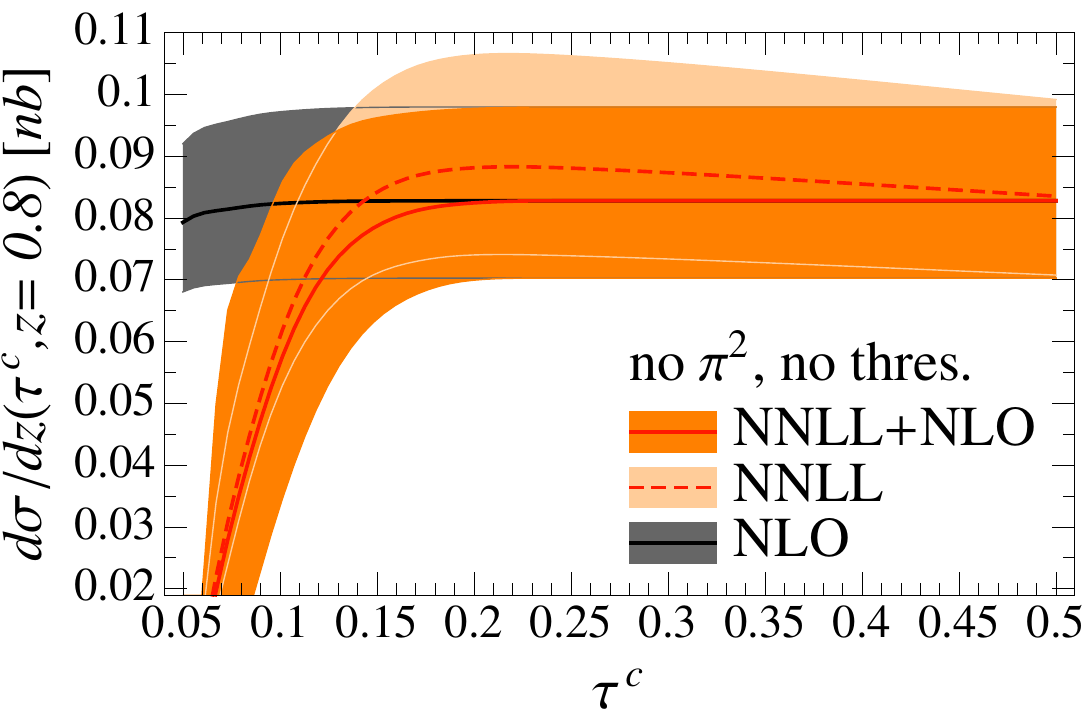}
\caption{The effect of the inclusion of the NLO nonsingular contribution is illustrated for $z=0.5$ (left panel) and $z=0.8$ (right panel), with $\pi^2$ and threshold resummations switched off.}
\label{fig:si_tau_ns}
\end{figure}

In \figs{si_tau}{si_tau_ns} the effects of the thrust cut and our resummations are illustrated for fixed values of $z$. As the cut becomes stronger (i.e. for smaller $\tau^\cut$) the cross section gets reduced, and by a larger amount in the case of our resummed calculation. For $\tau^\cut \lesssim 0.2$ our resummed result starts to differ from the leading order (LO) and NLO results, showing the effect of the resummation of double logarithms of $\tau^\cut$ and the leading nonperturbative correction. 

The nonsingular NLO correction is singled out in \fig{si_tau_ns}, where we switch off both the threshold resummation and the resummation of $\pi^2$-terms in the hard factor $H$ of \eq{Cj}. The nonsingular contribution is not large, but it causes our resummed calculation to merge with the NLO in the $\tau^\cut \to 0.5$ region where resummation is unimportant, and removes an unphysical decrease of the integrated cross section for large values of $\tau^\cut$. When we include threshold and $\pi^2$ resummation, we should also include them in the nonsingular contribution to the cross section, to preserve the above properties. Following ref.~\cite{Berger:2010xi}, in our analysis we estimate the effect of $\pi^2$ resummation in the nonsingular piece (which is formally of higher order in $\al_s$) by simply multiplying it by the corresponding evolution factor of the hard function $U_H(Q^2,\mu_H,|\mu_H|)$, given in eq.~(C.7) of ref.~\cite{Jain:2011xz}. We have not attempted to resum the threshold logarithms in the nonsingular contribution, so our calculation does not reliably describe the region where \emph{both} $\tau$ and $z$ are large.

\begin{figure}[t]
\centering
\includegraphics[width=0.50\textwidth]{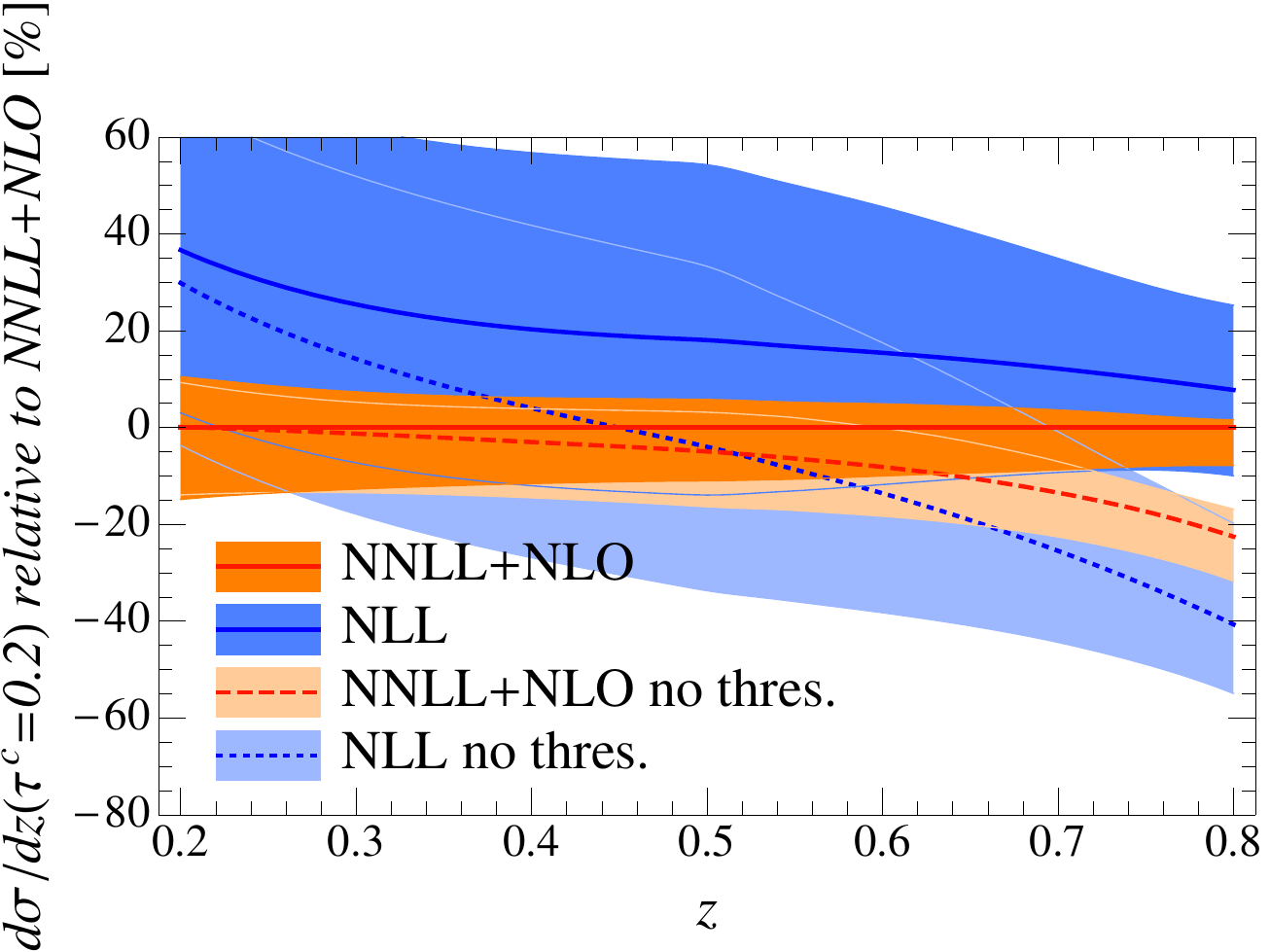}
 \caption{Comparison of NLL and NNLL+NLO with/without threshold resummation, for $\tau^\cut=0.2$.}
\label{fig:si_thr}
\end{figure}
\begin{figure}
\centering
\includegraphics[width=0.475\textwidth]{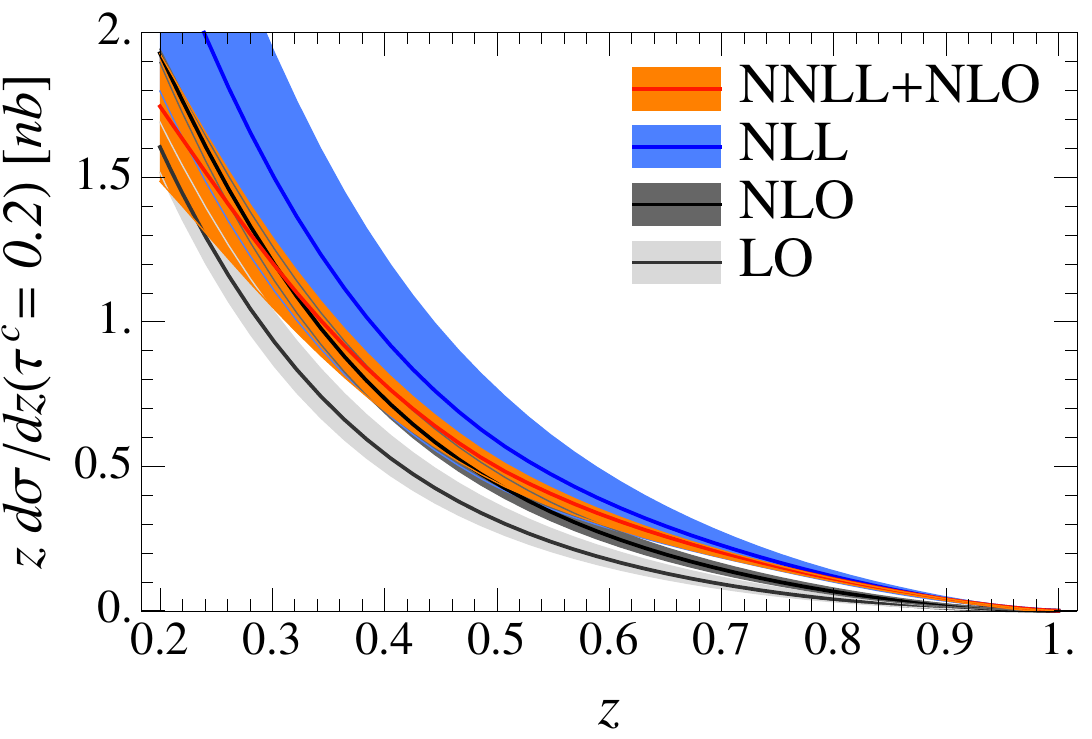} \quad
\includegraphics[width=0.48\textwidth]{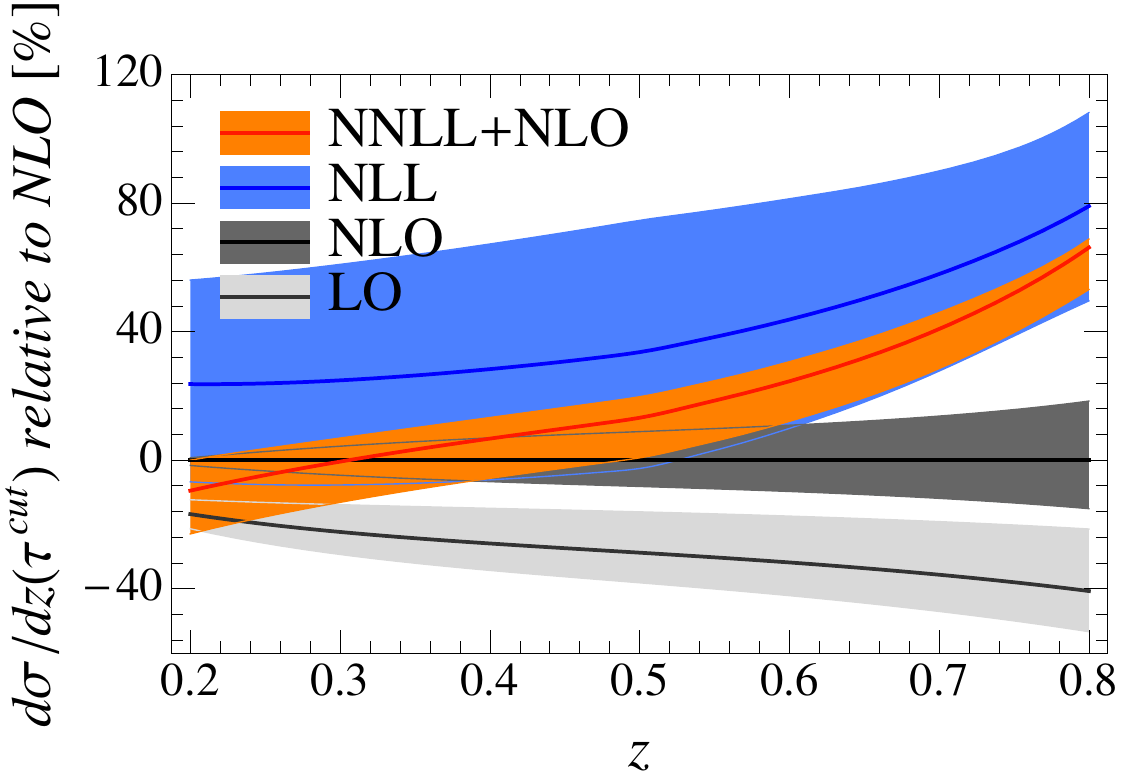}
\caption{The effect of the thrust cut $\tau^c =0.2$ on the $z$-spectrum for $e^+ e^- \to \text{dijet} + \pi^+$  with $Q=10.58$ GeV at LO, NLO, NLL and NNLL. The right panel shows the curves and bands in the left panel relative to the central curve for the NLO result.}
\label{fig:si_z}
\end{figure}
Next, we look at the effect of the thrust cut on the $z$ spectrum (``the shape of the fragmentation function"), which is shown in \figs{si_thr}{si_z} for the default value $\tau^c = 0.2$. In \fig{si_thr} we compare our resummed result at next-to-leading logarithmic order (NLL) and NNLL+NLO with and without threshold resummation.  
Comparing the two cases, our plot shows no difference concerning the quality of the convergence patterns and the size of the uncertainty bands. 
As was shown in ref.~\cite{Procura:2011aq}, the resummation of threshold logarithms is important for $z \gtrsim 0.5$. Since this resummation reduces the already small jet scale, nonperturbative corrections are more sizable and the improvement from threshold resummation is marginal, at variance with the dramatic effect in ref.~\cite{Procura:2011aq}.

In \fig{si_z} we compare our LO, NLL and NNLL+NLO curves with the NLO result, which is independent of $\tau^c$. The differences between the shape in $z$ depend on both the thrust cut and the order in (resummed) perturbation theory. As is clear from \fig{moneyplot}, we also find correlations between $\tau^\cut$ and $z$. 

\section{Testing $\tau^\cut$-$z$ correlations in Belle data}
\label{sec:prop}

In \fig{moneyplot} and \sec{num} we showed that the thrust cut has a sizable, calculable effect on the spectrum of the fragmentation variable $z$. Our framework allows one to easily check this 
in the Belle data. 

For a first quantitative analysis, we propose working in moment space, where \eq{factfirst} simplifies to
\begin{align} \label{eq:mom}
  \int_0^1\! \df z\, z^{N-1}\, \frac{\df \si^h}{\df z}(\tau^\cut) &= 
 \sum_j \tilde D_j^h(N,\mu) \,\tilde C_j(\tau^\cut, N, Q, \mu)
\,,\end{align}
with
\begin{align}
 \tilde D_j^h(N,\mu) &=   \int_0^1\! \df x\, x^{N-1}\, D_j^h(x,\mu)
\,.\end{align}
At variance with \eq{factfirst}, we use the notation $j= q, g$ by defining
\begin{equation} \label{eq:Cq}
  \tilde D_q^h = Q_u^2 \big(D_u^h+D_{\bar u}^h+D_c^h+D_{\bar c}^h\big) + 
  Q_d^2 \big(D_d^h+D_{\bar d}^h+D_s^h+D_{\bar s}^h\big)
  \,, \quad
  \tilde C_q = \tilde C_u/Q_u^2 \approx \tilde C_d/Q_d^2
\,,\end{equation}
with $Q_u=2/3$ and $Q_d=-1/3$ the electric charges of the up- and down-type quarks. 
In \app{coeff} we collect our numerical values for $\tilde C_j(\tau^\cut, N, Q = 10.58 \,\text{GeV}, \mu = 1 \,\text{GeV})$ in nanobarns for $j= q, g$, $\tau^\cut = 0.15,\,0.16,\dots,\,0.2$ and $N=3,4,\dots, 8$, at NNLL+NLO. We use $\alpha_s(m_Z)=0.118$ and three-loop running~\footnote{To avoid crossing flavor thresholds in our calculation, we choose to move the charm and bottom thresholds outside the range of scales that we work in. NNLL resummations require the running of $\alpha_s$ at three loops, see for example refs.~\cite{Jain:2011xz,Abbate:2010xh}.}, and employ two-loop splitting functions~\cite{Furmanski:1980cm,Curci:1980uw,Floratos:1981hs} to evolve to $\mu=1\,\text{GeV}$. We provide a different set of coefficients for each of the independent scale variations in \app{scales}, which allows one to take correlations between the uncertainties into account.

Apart from the different electric charges for the up- and down-type quarks, the difference between $\tilde C_u$ and $\tilde C_d$ is less than one percent. Thus the thrust cut 
is not sensitive to flavor-separated FFs 
since it involves only the combination in \eq{Cq}. In contrast, varying the thrust cut allows one to be sensitive to the separate gluon contribution. However, at the scale $\mu=1\, \text{GeV}$, which is the input scale for fragmentation functions, this contribution to the cross section is small and dominated by uncertainties. 
\begin{figure}
\centering
\includegraphics[width=0.48\textwidth]{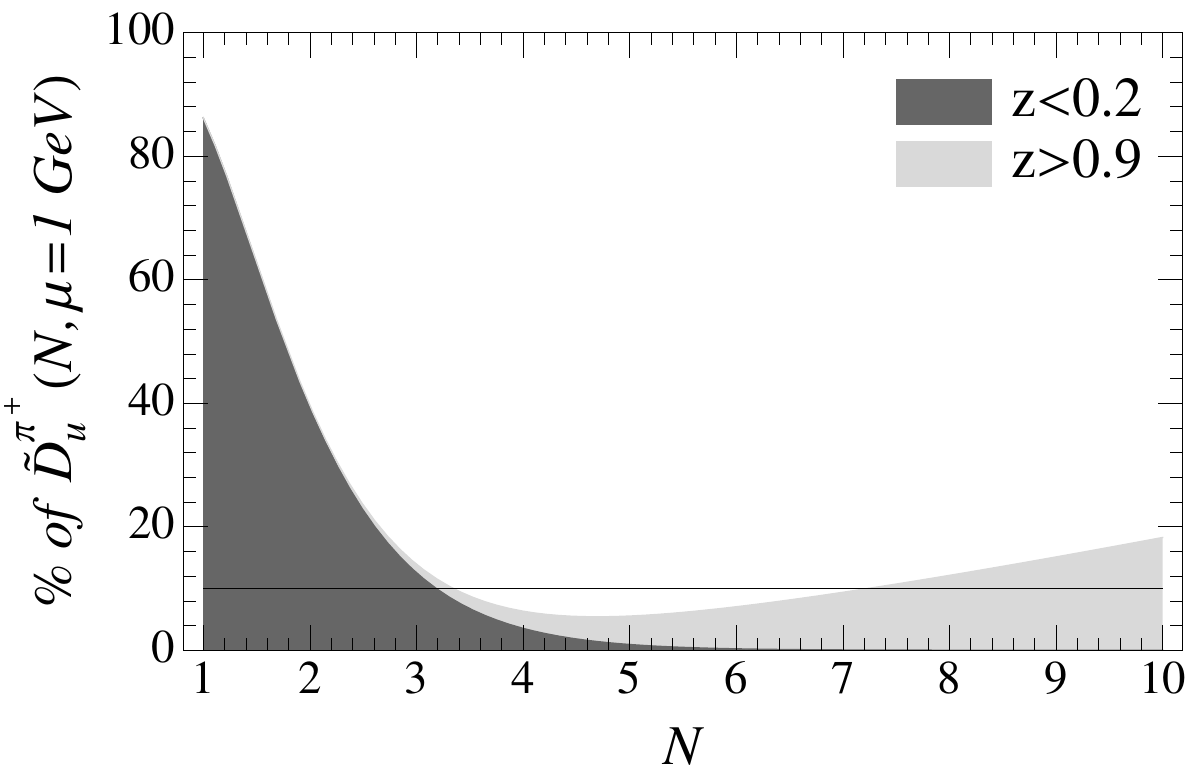}
\quad
\includegraphics[width=0.48\textwidth]{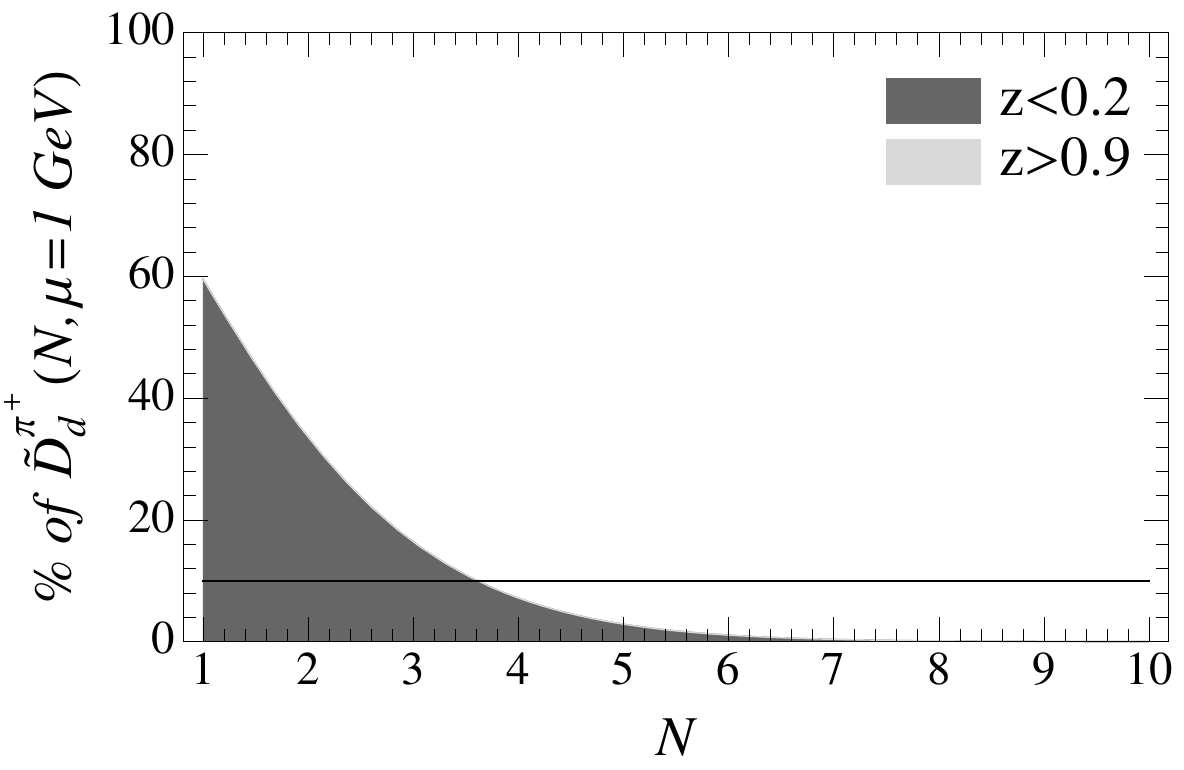} 
\caption{The composition of the $N$th moment of $D_u^{\pi^+}$ (left panel) and $D_d^{\pi^+}$ (right panel). Using different colors we show the contribution from the problematic regions $z<0.2$ and $z>0.9$. The contribution from $z>0.9$ for $d$ quarks is very small and hardly visible.}
\label{fig:mom}
\end{figure}

Taking as input the moments $\tilde D_q(N, \mu = 1\, \text{GeV})$ from the available phenomenological parameterization of the FFs together with the numbers in table \ref{tab:Ci}, \eq{mom} can be directly compared with the Belle spectra.
 We stress that our formalism can be more stringently tested by using data at different $\tau^\cut$.

Note that Mellin moment methods have been used in analyzing fragmentation functions before~\cite{deFlorian:2007aj,deFlorian:2007hc}.
Experimentally, using moments has the disadvantage that they can be sensitive to the problematic $z\lesssim0.2$ region, and the $z\gtrsim0.9$ region, where the experimental uncertainties are large~\cite{Leitgab:2011zz}. The relative contribution of these regions to the moments of $D_u^{\pi^+}$ and $D_d^{\pi^+}$ is shown in \fig{mom}. This is why we restricted ourselves here to moments in the range $3 \leq N \leq 8$. In addition, the thrust cut should neither be too soft (contamination with $b$ quarks) nor too strong (reduced signal). As we explained in \sec{calc}, the requirements for our calculation to be under theoretical control are similar.

\section{Conclusions}
\label{sec:conc}

We have studied the cross section for fragmentation with a thrust cut in a formalism that can easily be tested in high-statistics on-resonance B-factory data. The shape of the cross section in $z$ is altered beyond LO and includes a $\tau$-dependent contribution. The correlation between $\tau$ and $z$ found in our resummed calculation is consistent with \Pythia as well, but is much smaller at NLO. Our plots in \secs{intro}{num} and the discussion in \sec{prop} provide the starting point for a quantitative check of this effects and of our resummations. We emphasize that binning the data in $\tau^\cut$ allows one to study the correlations in $\tau^\cut$ and $z$, and leads to more stringent tests.

\begin{acknowledgments}
We thank M. Grosse Perdekamp for helpful discussions. M.P.~acknowledges support  by the ``Innovations- und Kooperationsprojekt C-13'' of the Schweizerische Universit\"atskonferenz SUK/CRUS and by the Swiss National Science Foundation. W.W.~is supported by DOE grant DE-FG02-90ER40546. 
\end{acknowledgments}

\appendix

\section{Nonsingular contribution at NLO}
\label{app:nonsingular}

We obtain the NLO nonsingular correction to the cross section for $e^+ e^- \to \text{dijet} + h$, by subtracting the singular $\ln^m \tau^\cut$ terms from the full NLO calculation. The singular NLO can directly be obtained from our resummed calculation in \eq{Cj}, so we start by calculating the full NLO cross section. 

The easiest way to obtain the fragmentation cross section with a thrust cut at NLO is
\begin{equation} \label{eq:si_z_tau}
  \frac{\df \si_h}{\df z}(\tau \leq \tau^\cut) = \frac{\df \si_h}{\df z} - \frac{\df \si_h}{\df z}(\tau \geq \tau^\cut)
\,.\end{equation}
The first term on the right-hand side is the cross section without a thrust cut, 
\begin{align} \label{eq:si_z_fact}
  \frac{\df \si_h}{\df z} &= \sum_{i=q,\bar q,g} \int_z^1\! \frac{\df x}{x}\, \frac{\df \hat \si_i}{\df x}\, D_i^h\Big(\frac{z}{x}\Big)
\,,\end{align}
which was calculated at NLO a long time ago~\cite{Altarelli:1979kv,Curci:1980uw,Floratos:1981hs}
\begin{align}
  \frac{\df \hat \si_{q,R}^\one}{\df z} &
  = \si^\zero\, \frac{\al_s C_F}{\pi} 
  \Big\{
 - \frac{1}{2} P_{qq}(z) \ln \frac{\mu^2}{q^2} + 
  (1+z^2) \Big[\frac{1}{2}\Big(\frac{\ln(1-z)}{1-z}\Big)_+ + \frac{\ln z}{1-z}\Big]  - \frac{3}{4} \frac{1}{(1-z)}_+ +
  \nn \\ & \quad
  \de(1-z) \Big(\frac{\pi^2}{3} - \frac{9}{4}\Big) - \frac{3}{4} z + \frac{5}{4} \Big\}
  \,, \nn \\
  \frac{\df \hat \si_g^\one}{\df z} &
  = \si^\zero \, \frac{\al_s C_F}{\pi}\,
   P_{gq}(z) \Big[ - \ln \frac{\mu^2}{q^2} + \ln(1-z)+2 \ln z\Big]  
\,.\end{align}
The tree-level cross section $\si^\zero = 4\pi \al^2 N_c/(3q^2)$ and the splitting functions are given by~\cite{Altarelli:1977zs}
\begin{align} \label{eq:split}
  P_{qq}(z) = \Big(\frac{1+z^2}{1-z}\Big)_+
  = \frac{1+z^2}{(1-z)}_+ + \frac{3}{2}\, \de(1-z)
  \,, \qquad
  P_{gq}(z) = \theta(1-z) \,\frac{1+(1-z)^2}{z}
  \,.
\end{align}
The second term in \eq{si_z_tau} is described by a factorization theorem very similar to \eq{si_z_fact}
\begin{align} 
  \frac{\df \si_h}{\df z}(\tau \geq \tau^\cut) = \sum_{i=q,\bar q,g} \int_z^1\! \frac{\df x}{x}\, \frac{\df \hat \si_i}{\df x}(\tau \geq \tau^\cut)\, D_i^h\Big(\frac{z}{x}\Big)
\,.\end{align}
At NLO the thrust cut removes any regions of phase-space involving divergences and only the real radiation contributes, 
\begin{align}
  \frac{\df \hat \si_q}{\df z}(\tau \geq \tau^\cut) &
  = \si^\zero\, \frac{\al_s C_F}{2\pi}\,
\theta(1-z-\tau^\cut) \theta(z-2\tau^\cut)
 \int_{1-z+\tau^\cut}^{1-\tau^\cut}\!\! \df x\,\frac{z^2 + x^2}{(1-z)(1-x)} 
\nn \\ &
  = \si^\zero \frac{\al_s C_F}{\pi} \, \theta(1-z-\tau^\cut) \theta(z-2\tau^\cut)
  \Big[\frac{1}{2} P_{qq}(z) \ln \frac{z-\tau^\cut}{\tau^\cut} + \frac{z^2-2z(\tau^\cut+2)+8\tau^\cut}{4(1-z)}  \Big]
 \,, \nn \\
  \frac{\df \hat \si_g}{\df z}(\tau \geq \tau^\cut) &
  = \si^\zero\, \frac{\al_s C_F}{2\pi} 
\theta(1-z-\tau^\cut) \theta(z-2\tau^\cut)
 \int_{1-z+\tau^\cut}^{1-\tau^\cut}\!\! \df x\,\frac{(2-x-z)^2 + x^2}{(x+z-1)(1-x)} 
\nn \\ &
  = \si^\zero\, \frac{\al_s C_F}{\pi} \, \theta(1-z-\tau^\cut) \theta(z-2\tau^\cut)
  \Big[P_{gq}(z) \ln \frac{z-\tau^\cut}{\tau^\cut} - z+2\tau^\cut  \Big]
 \,.\end{align}

%
%
\begin{figure}[t]
\centering
\includegraphics[width=0.48\textwidth]{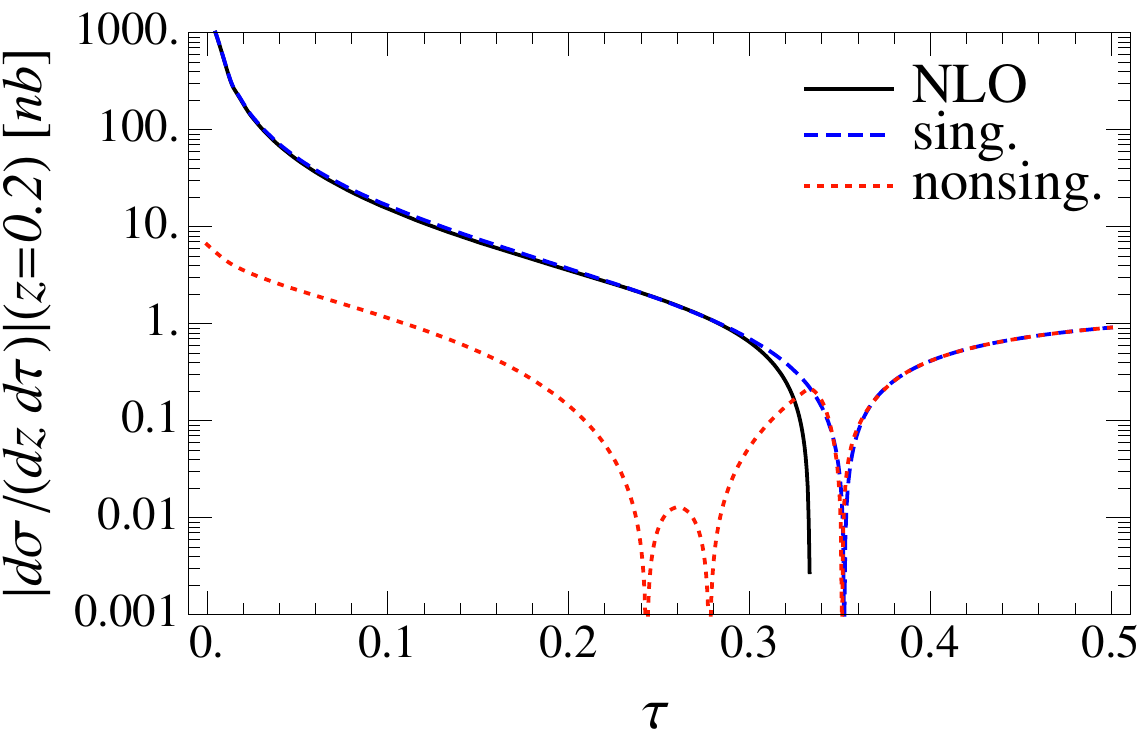} \quad
\includegraphics[width=0.48\textwidth]{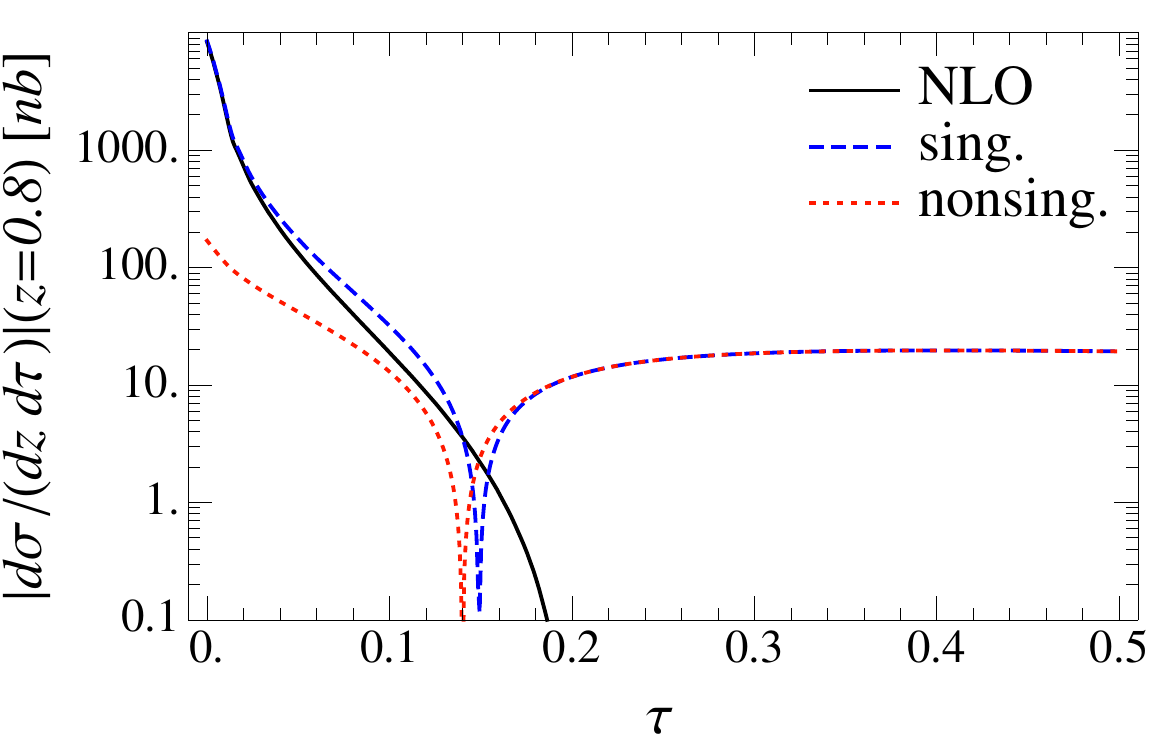}
\caption{The cross section for $e^+ e^- \to \text{dijet} + \pi^+$ at $Q=10.58$ GeV, differential in the momentum fraction $z$ and thrust $\tau$, and separated into its singular and nonsingular contribution. Since these separate pieces can be negative, absolute values are plotted.}
\label{fig:nonsing}
\end{figure}

We now arrive at the nonsingular cross section by subtracting the singular cross section, which can directly be obtained from our resummed calculation in \eq{Cj} by choosing the renormalization scale of all the objects equal to $\mu$. In \fig{nonsing} the NLO cross section for $z=0.2$ and $z=0.8$ is shown separated into its singular and nonsingular contribution. A first cross-check of our calculation is provided by the fact that the nonsingular piece is finite in the $\tau \to 0$ limit, whereas both the full NLO and the singular piece diverge. The contribution of the nonsingular to the total cross section is suppressed for small $\tau$ by $\ord{\tau}$. However, for larger values of $\tau$ its contribution becomes more important, and above the NLO kinematic threshold $\tau \geq \min\{1/3, 1-z\}$ the total cross section vanishes and the singular and nonsingular exactly cancel each other. In this region the inclusion of the nonsingular is essential to avoid a negative differential cross section in $\tau$ or, equivalently, an unphysical turn-over in the cross section integrated up to $\tau \leq \tau^\cut$.

\section{Choice of running scales}
\label{app:scales}

The resummation of logarithms of $\tau$ and $1-z$, as well as large $\pi^2$ terms, is accomplished by evaluating $H$, $J$, $\cJ$ and $S$ in \eq{Cj} at their natural scales $\mu_H$, $\mu_J$, $\mu_\cJ$, $\mu_S$, and using their respective RGEs to evolve them to a common scale $\mu$. Having made some improvements compared to our earlier work ref.~\cite{Jain:2011xz}, we discuss this in some detail here.

The hard function $H$ is the square of a time-like form factor and contains large $\pi^2$-terms at $\mu_H = Q$ arising from $\ln^2(-\img Q/\mu_H)$, see \eq{hard}. These can be resummed through the complex scale setting $\mu_H=-\img Q$~\cite{Parisi:1979xd, Sterman:1986aj, Magnea:1990zb, Eynck:2003fn}. Following ref.~\cite{Abbate:2010xh}, we note that there are three distinct kinematic regions where the resummation of the logarithms of $\tau$ must be handled differently:
\begin{align}
\text{1)}& \quad
\mu_H \simeq -\img Q\,,\qquad \mu_J \simeq \sqrt{\lqcd Q}\,, \qquad \mu_S = \lqcd
\,, \hspace{20ex}\nn\\
\text{2)}& \quad
\mu_H \simeq -\img Q\,, \qquad \mu_J \simeq \sqrt{\tau} Q\,, \qquad \mu_S \simeq \tau Q
\,,\nn\\
\text{3)}& \quad
\img \mu_H = \mu_J = \mu_S \simeq Q
\,.\nn\end{align}
Region 1) corresponds to $\tau_\cut \sim \lqcd/Q$ and requires a completely nonperturbative soft function. Physically this corresponds to the limit where there are only a few hadrons in each hemisphere and jet. This region is not of interest for our application.
Region 2) corresponds to $\tau \ll 1$ but large enough so that $\tau Q$ is still perturbative. This region is of interest to us and corresponds to factorizable contributions in the first term of eq. (\ref{eq:Cj}). Region 3) corresponds to large values of $\tau$ and to the second term in \eq{Cj}.
The resummation of the threshold logarithms of $1-z$ sets the scale $\mu_\cJ$ different from $\mu_J$ for the \emph{diagonal} terms $\cJ_{qq}$~\cite{Procura:2011aq}. In addition, the measurement of $z$ affects the upper bound $\tau \leq \tau_{\max} = \min\{0.5,1-z\}$, implying that region 3) sets in at smaller values of $\tau$ for $z > 0.5$. 
Taking this into account, a smooth transition between the three regions is given by
\begin{align} \label{eq:scales}
  \mu_H &= -\img\, e_H\, Q
  \,, \nn \\
  \mu_J(\tau,z) & =
  \Big[1 + e_J\, \theta(r_3 \tau_{\max}-\tau) \Big(1 - \frac{\tau}{r_3 \tau_{\max}}\Big)^2\,\Big]
  \sqrt{|\mu_H|\, \mu_\mathrm{run}(\tau/\tau_{\max},z,|\mu_H|)}
  \,, \nn \\
  \mu_\cJ(\tau,z) & = \sqrt{1-z} \ \mu_J(\tau,z)
  \,, \nn \\
  \mu_S(\tau,z) & = \Big[1 + e_S\, \theta(r_3 \tau_{\max}-\tau) \Big(1 - \frac{\tau}{r_3 \tau_{\max}}\Big)^2\,\Big] \mu_\mathrm{run}(\tau/\tau_{\max},z,|\mu_H|)
  \,,
\end{align}
where scale uncertainties will be estimated by varying $e_H$, $e_J$ and $e_S$ as shown in table~\ref{tab:vars}. In addition to the explicit $z$-dependence in $\mu_\cJ$, there is an implicit dependence through $\tau_{\max}$. 
For the profile $\mu_\mathrm{run}(\tau/\tau_{\max},z,|\mu_H|)$ we use a combination of two quadratic functions and a linear function as in refs.~\cite{Abbate:2010xh,Berger:2010xi},
\begin{align} \label{eq:murun}
& \mu_\mathrm{run}(r,z,|\mu_H|) =
\begin{cases}
\mu_0 + a\,r^2/r_1 & r \leq r_1
\,,\\
2a\, r + b & r_1 \leq r \leq r_2
\,,\\
|\mu_H| - a (r-r_3)^2/(r_3 - r_2) & r_2 \leq r \leq r_3 
\,,\\
|\mu_H| & r > r_3
\,,\end{cases}
\nn \\
&
a= \frac{\mu_0-|\mu_H|}{r_1 -r_2 -r_3}
\,, \qquad
b = \frac{|\mu_H| r_1 - \mu_0 (r_2 + r_3)}{r_1-r_2-r_3}
\,.\end{align}
The expressions for $a$ and $b$ follow from demanding that $\mu_\mathrm{run}$ is continuous and has a continuous derivative. The value of $\mu_0$ determines the scales at $\tau = 0$, while $r_{1,2,3}$ determine the transition between the regions discussed above. For $\tau > r_3 \tau_{\max}$, our choice for $\mu_\mathrm{run}$ ensures that the resummation of logarithms of $\tau$ turns off (but $\pi^2$ resummation and threshold resummation are still present). 

For the cross section integrated up to $\tau^\cut$, we use the same scale choices as in \eq{scales} with $\tau \to \tau^\cut$. In moment space, $\tau_{\max} = 0.5$ and the equation for $\mu_\cJ$ is modified to
\begin{equation}
  \mu_\cJ(\tau,N) = \frac{\mu_J(\tau)}{\sqrt{N} e^{\ga_E/2}}
\,.\end{equation}
We complete our description by specifying our choice for the parameters of the profile function in \eq{murun}. Our central value corresponds to:
\begin{equation} \label{eq:profile}
e_H = 1
\,,\quad
e_J = e_S = 0
\,,\quad
\mu_0 = 2 \GeV
\,,\quad
r_1 = \frac{4 \GeV}{Q}
\,,\quad
r_2 = 0.5
\,,\quad
r_3 = 1
\,.\end{equation}
As mentioned before, we estimate the perturbative uncertainties by a combination of separate scale variations, which are listed in table~\ref{tab:vars}. 
In our cross section plots we add in quadrature the upward and downward variations, which we, respectively, determine as the largest and smallest value of the cross section obtained by looking at the pairs of variations for the hard scale, the jet scale and the soft scale. 
In the tables in \app{coeff} we give results for all the separate scale variations, allowing one to take correlations in the uncertainties into account.
\begin{table}
\centering
\begin{tabular}{|l|rrr|l|}
\hline
Variation & $e_H$ & $e_J$ & $e_S$ & Interpretation\\
\hline
1& 1 & 0 & 0 & Central value\\ 
2& 2 & 0 & 0 & \multirow{2}{*}{Hard scale variation} \\
3 & 0.5 & 0 & 0 & \\
4 & 1 & 0.5 & 0 & \multirow{2}{*}{Jet scale variation} \\
5 & 1 & -0.5 & 0 & \\
6 & 1 & 0 & 0.5 & \multirow{2}{*}{Soft scale variation} \\
7 & 1 & 0 & -0.5 & \\
\hline
\end{tabular}
\caption{The scale variations in terms of the parameters in \eq{scales}.}
\label{tab:vars}
\end{table}

\section{Perturbative coefficients $C_j$ in moment space}
\label{app:coeff}

In this appendix we give numerical results for the coefficients in \eq{mom} at NNLL+NLO. Table \ref{tab:Ci}  contains the results for $\tilde C_q$ and $\tilde C_g$. For each of the independent scale variations in table \ref{tab:vars} there is a separate set of $\tilde C_j$, labelled by the number in the first column. 

\newpage 
\begin{table*}
\vspace{-2ex}
\centering
$\tilde C_q:$ \quad
\begin{tabular}{|ll|rrrrrr|}
\hline
 var & $\tau^\cut$ & 3 & 4 & 5 & 6 & 7 & 8\\ 
 \hline 
1 & 0.15 & 1.69 & 1.70 & 1.71 & 1.72 & 1.74 & 1.75 \\ 
 & 0.16 & 1.77 & 1.76 & 1.76 & 1.76 & 1.77 & 1.78 \\ 
 & 0.17 & 1.84 & 1.81 & 1.80 & 1.79 & 1.79 & 1.79 \\ 
 & 0.18 & 1.89 & 1.86 & 1.83 & 1.82 & 1.81 & 1.80 \\ 
 & 0.19 & 1.94 & 1.89 & 1.86 & 1.84 & 1.82 & 1.81 \\ 
 & 0.20 & 1.99 & 1.93 & 1.88 & 1.85 & 1.83 & 1.81 \\ 
 \hline 
2 & 0.15 & 1.73 & 1.73 & 1.73 & 1.74 & 1.76 & 1.77 \\ 
 & 0.16 & 1.82 & 1.80 & 1.79 & 1.80 & 1.80 & 1.81 \\ 
 & 0.17 & 1.90 & 1.86 & 1.85 & 1.84 & 1.83 & 1.84 \\ 
 & 0.18 & 1.96 & 1.92 & 1.89 & 1.87 & 1.86 & 1.85 \\ 
 & 0.19 & 2.02 & 1.96 & 1.92 & 1.89 & 1.88 & 1.87 \\ 
 & 0.20 & 2.06 & 1.99 & 1.95 & 1.91 & 1.89 & 1.88 \\ 
 \hline 
3 & 0.15 & 1.68 & 1.69 & 1.69 & 1.70 & 1.70 & 1.70 \\ 
 & 0.16 & 1.74 & 1.73 & 1.72 & 1.71 & 1.70 & 1.69 \\ 
 & 0.17 & 1.79 & 1.76 & 1.74 & 1.72 & 1.71 & 1.69 \\ 
 & 0.18 & 1.83 & 1.79 & 1.76 & 1.73 & 1.71 & 1.68 \\ 
 & 0.19 & 1.87 & 1.82 & 1.77 & 1.74 & 1.70 & 1.67 \\ 
 & 0.20 & 1.90 & 1.84 & 1.78 & 1.74 & 1.70 & 1.67 \\ 
 \hline 
4 & 0.15 & 1.62 & 1.62 & 1.62 & 1.63 & 1.65 & 1.66 \\ 
 & 0.16 & 1.71 & 1.69 & 1.69 & 1.69 & 1.69 & 1.70 \\ 
 & 0.17 & 1.78 & 1.75 & 1.74 & 1.73 & 1.73 & 1.74 \\ 
 & 0.18 & 1.84 & 1.80 & 1.78 & 1.77 & 1.76 & 1.76 \\ 
 & 0.19 & 1.90 & 1.85 & 1.82 & 1.80 & 1.78 & 1.77 \\ 
 & 0.20 & 1.95 & 1.89 & 1.85 & 1.82 & 1.80 & 1.79 \\ 
 \hline 
5 & 0.15 & 1.83 & 1.85 & 1.86 & 1.88 & 1.88 & 1.89 \\ 
 & 0.16 & 1.89 & 1.89 & 1.89 & 1.89 & 1.88 & 1.88 \\ 
 & 0.17 & 1.94 & 1.92 & 1.90 & 1.89 & 1.88 & 1.87 \\ 
 & 0.18 & 1.98 & 1.94 & 1.92 & 1.90 & 1.88 & 1.87 \\ 
 & 0.19 & 2.01 & 1.97 & 1.93 & 1.90 & 1.88 & 1.86 \\ 
 & 0.20 & 2.05 & 1.99 & 1.94 & 1.91 & 1.88 & 1.85 \\ 
 \hline 
6 & 0.15 & 1.84 & 1.85 & 1.85 & 1.86 & 1.87 & 1.88 \\ 
 & 0.16 & 1.91 & 1.89 & 1.89 & 1.88 & 1.88 & 1.89 \\ 
 & 0.17 & 1.96 & 1.93 & 1.91 & 1.90 & 1.89 & 1.88 \\ 
 & 0.18 & 2.00 & 1.96 & 1.93 & 1.91 & 1.89 & 1.88 \\ 
 & 0.19 & 2.04 & 1.98 & 1.94 & 1.92 & 1.89 & 1.88 \\ 
 & 0.20 & 2.07 & 2.00 & 1.96 & 1.92 & 1.89 & 1.87 \\
 \hline
7 & 0.15 & 1.48 & 1.50 & 1.52 & 1.54 & 1.57 & 1.59 \\ 
 & 0.16 & 1.59 & 1.60 & 1.60 & 1.62 & 1.63 & 1.65 \\ 
 & 0.17 & 1.69 & 1.68 & 1.67 & 1.68 & 1.68 & 1.69 \\ 
 & 0.18 & 1.77 & 1.74 & 1.73 & 1.72 & 1.72 & 1.72 \\ 
 & 0.19 & 1.83 & 1.80 & 1.77 & 1.76 & 1.75 & 1.75 \\ 
 & 0.20 & 1.89 & 1.84 & 1.81 & 1.79 & 1.77 & 1.76 \\ 
 \hline 
\end{tabular} \qquad \quad
$\tilde C_g:$ \quad
\begin{tabular}{|ll|rrrrrr|}
\hline
 var & $\tau^\cut$ & 3 & 4 & 5 & 6 & 7 & 8\\ 
 \hline
1 & 0.15 & 0.257 & 0.167 & 0.112 & 0.079 & 0.057 & 0.042 \\ 
 & 0.16 & 0.291 & 0.185 & 0.123 & 0.086 & 0.063 & 0.046 \\ 
 & 0.17 & 0.323 & 0.201 & 0.133 & 0.094 & 0.068 & 0.051 \\ 
 & 0.18 & 0.353 & 0.217 & 0.143 & 0.101 & 0.074 & 0.055 \\ 
 & 0.19 & 0.382 & 0.232 & 0.153 & 0.107 & 0.079 & 0.060 \\ 
 & 0.20 & 0.409 & 0.246 & 0.161 & 0.113 & 0.083 & 0.063 \\ 
 \hline 
2 & 0.15 & 0.441 & 0.284 & 0.198 & 0.148 & 0.116 & 0.094 \\ 
 & 0.16 & 0.486 & 0.307 & 0.212 & 0.157 & 0.123 & 0.099 \\ 
 & 0.17 & 0.526 & 0.328 & 0.225 & 0.166 & 0.129 & 0.104 \\ 
 & 0.18 & 0.564 & 0.346 & 0.236 & 0.174 & 0.134 & 0.108 \\ 
 & 0.19 & 0.598 & 0.364 & 0.246 & 0.180 & 0.139 & 0.111 \\ 
 & 0.20 & 0.630 & 0.379 & 0.255 & 0.186 & 0.143 & 0.115 \\ 
 \hline 
3 & 0.15 & 0.060 & 0.019 & -0.011 & -0.030 & -0.043 & -0.053 \\ 
 & 0.16 & 0.083 & 0.031 & -0.004 & -0.025 & -0.039 & -0.049 \\ 
 & 0.17 & 0.106 & 0.042 & 0.004 & -0.019 & -0.034 & -0.044 \\ 
 & 0.18 & 0.128 & 0.054 & 0.011 & -0.013 & -0.029 & -0.040 \\ 
 & 0.19 & 0.149 & 0.065 & 0.019 & -0.007 & -0.024 & -0.035 \\ 
 & 0.20 & 0.169 & 0.076 & 0.026 & -0.001 & -0.019 & -0.030 \\ 
 \hline 
4 & 0.15 & 0.374 & 0.235 & 0.161 & 0.118 & 0.091 & 0.072 \\ 
 & 0.16 & 0.404 & 0.250 & 0.170 & 0.124 & 0.095 & 0.075 \\ 
 & 0.17 & 0.431 & 0.265 & 0.179 & 0.130 & 0.099 & 0.078 \\ 
 & 0.18 & 0.457 & 0.278 & 0.186 & 0.135 & 0.103 & 0.081 \\ 
 & 0.19 & 0.481 & 0.289 & 0.193 & 0.139 & 0.106 & 0.083 \\ 
 & 0.20 & 0.503 & 0.300 & 0.200 & 0.144 & 0.109 & 0.085 \\ 
 \hline 
5 & 0.15 & 0.086 & 0.045 & 0.012 & -0.010 & -0.025 & -0.036 \\ 
 & 0.16 & 0.131 & 0.073 & 0.032 & 0.007 & -0.010 & -0.022 \\ 
 & 0.17 & 0.174 & 0.099 & 0.051 & 0.023 & 0.004 & -0.008 \\ 
 & 0.18 & 0.215 & 0.123 & 0.069 & 0.037 & 0.017 & 0.003 \\ 
 & 0.19 & 0.254 & 0.146 & 0.086 & 0.051 & 0.029 & 0.014 \\ 
 & 0.20 & 0.291 & 0.168 & 0.101 & 0.063 & 0.039 & 0.023 \\ 
 \hline 
6 & 0.15 & 0.272 & 0.178 & 0.120 & 0.084 & 0.060 & 0.044 \\ 
 & 0.16 & 0.306 & 0.195 & 0.130 & 0.091 & 0.066 & 0.048 \\ 
 & 0.17 & 0.338 & 0.211 & 0.140 & 0.098 & 0.071 & 0.053 \\ 
 & 0.18 & 0.368 & 0.227 & 0.149 & 0.104 & 0.076 & 0.057 \\ 
 & 0.19 & 0.396 & 0.241 & 0.158 & 0.111 & 0.081 & 0.061 \\ 
 & 0.20 & 0.422 & 0.254 & 0.166 & 0.117 & 0.086 & 0.065 \\ 
 \hline
7 & 0.15 & 0.236 & 0.153 & 0.103 & 0.073 & 0.053 & 0.039 \\ 
 & 0.16 & 0.272 & 0.172 & 0.115 & 0.081 & 0.059 & 0.044 \\ 
 & 0.17 & 0.305 & 0.190 & 0.127 & 0.089 & 0.066 & 0.049 \\ 
 & 0.18 & 0.337 & 0.207 & 0.137 & 0.097 & 0.071 & 0.054 \\ 
 & 0.19 & 0.368 & 0.223 & 0.147 & 0.104 & 0.077 & 0.058 \\ 
 & 0.20 & 0.396 & 0.239 & 0.157 & 0.111 & 0.082 & 0.062 \\ 
 \hline
\end{tabular}
\caption{$\tilde C_i(\tau^\cut, N, Q = 10.58 \,\text{GeV}, \mu = 1 \,\text{GeV})$ in units of nb at NNLL+NLO, with scale variations. $\tilde C_q$ applies to both up and down-type quarks and was defined in \eq{Cq}.}
\label{tab:Ci}
\end{table*}

\newpage 
\bibliographystyle{physrev4}
\bibliography{../Fragmentation}


\end{document}